\documentclass[10pt,conference]{IEEEtran}

\usepackage{epsfig}
\usepackage{epsfig}
\newcommand{\ot}{{2 \choose 1}{\rm -OT}}
\begin{document}

\title{Oblivious Transfer and Quantum Channels}

\author{\authorblockN{Nicolas Gisin\authorrefmark{1}\qquad
Sandu Popescu\authorrefmark{2}\authorrefmark{3}\qquad
Valerio Scarani\authorrefmark{1}\qquad
Stefan Wolf\authorrefmark{4}\qquad
J\"urg Wullschleger\authorrefmark{4}}
\authorblockA{\authorrefmark{1}Group of Applied Physics,
University of Geneva,
10, rue de l'\'Ecole-de-M\'edecine,
CH-1211 Gen\`eve 4,
Switzerland.\\
E-mail: \{Nicolas.Gisin,Valerio.Scarani\}@physics.unige.ch}
\authorblockA{\authorrefmark{2}H.\ H.\ Wills Physics Laboratory, University of Bristol, Tyndall Avenue, Bristol BS8 1TL, U.K.}
\authorblockA{\authorrefmark{3}Hewlett-Packard Laboratories, Stoke Gifford, Bristol BS12 6QZ, U.K.\\ 
E-mail: S.Popescu@bristol.ac.uk}
\authorblockA{\authorrefmark{4}Computer Science Department,
ETH Z\"urich,
ETH Zentrum,
CH-8092 Z\"urich,
Switzerland.\\
E-mail: \{wolf,wjuerg\}@inf.ethz.ch}}

\maketitle

\begin{abstract}
We show that oblivious transfer can be seen as the classical analogue to 
a quantum channel in the same sense as non-local boxes are for maximally
entangled qubits.
\end{abstract}

\section{Introduction}

\subsection{Quantum Entanglement and Non-Locality}

One of the most interesting and surprising consequences of the laws of quantum physics 
is the phenomenon of {\em entanglement}. In 1935, Einstein, Podolsky, and Rosen~\cite{EPR} 
initiated a discussion on non-local behavior. Let us consider, for instance,
the following,  {\em maximally entangled\/} state, called {\em singlet\/} or, 
alternatively,  {\em EPR\/}
or {\em Bell state}: 
\begin{equation}\label{singlet}
|\psi^-\rangle=\frac{1}{\sqrt{2}}(|01\rangle-|10\rangle)\ .
\end{equation}

The state $|\psi^-\rangle$ has the property that when the two qubits
are measured in the same basis, the measurements lead to perfectly anti-correlated results|even 
when the two measurement events are spatially separated. The conclusion of~\cite{EPR}
was that quantum physics was incomplete in the sense that it should be augmented by certain 
{\em hidden parameters\/} determining the results of measurements. However, it was shown 
later by von Neumann~\cite{vN}, Gleason~\cite{gleason}, Specker~\cite{specker}, 
Jauch and Piron~\cite{jp}, 
Bell~\cite{Bell}, and Kochen and Specker~\cite{kospe}
that such hidden-parameter models fail to explain quantum-physical behavior in general.

\begin{figure}[h]
\begin{center}
\input{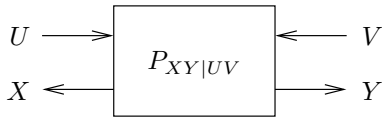}
\end{center}
\caption{A two-party information-theoretic primitive.}
\label{pxyuv}
\end{figure}

The behavior of a bipartite quantum state under measurements can be described by a 
conditional probability distribution $P_{XY|UV}$|also called two-party information-theoretic 
primitive|where $U$ and $V$ denote the chosen bases and $X$ and $Y$ the corresponding 
outcomes (see Figure~\ref{pxyuv}). 

\begin{figure}[h]
\begin{center}
\input{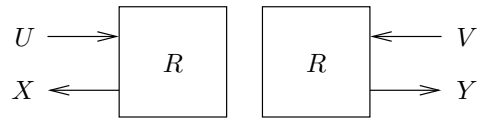}
\end{center}
\caption{A local system.}
\label{local}
\end{figure}

Bell~\cite{Bell} was the first to recognize that there exist pairs of measurement bases such that
the resulting behavior is not local, i.e., cannot be explained 
by shared classical information (see Figure~\ref{local}). He showed that there exist certain
inequalities (now called \emph{Bell inequalities}) that cannot be violated by any local system,
but which {\em are\/} violated when  an EPR pair is measured.

The  behavior of $|\psi^-\rangle$ is, hence, {\em non-local}.
It is important to note that  non-local behavior, although 
explainable classically  only by communication, does {\em not\/} allow for 
 signaling. In order to understand non-locality and 
its consequences better, Popescu and Rohrlich~\cite{poro} have introduced
a ``non-locality machine,'' called {\em non-local box\/} or {\em NL box\/}
for short.
The behavior of the NL box is inspired by the one of $|\psi^-\rangle$, but it is
 ``more non-local'' than the one 
of any quantum state. 
The (binary) inputs and outputs of the NL box are $(U,V)$ and $(X,Y)$, respectively, 
and $X$ and $Y$ are random bits satisfying 
\begin{eqnarray*}
{\rm Prob}[X=Y|(U,V)\ne (1,1)]& =& 1\ ,\\
{\rm Prob}[X=Y|(U,V)= (1,1)]& =& 0\ .
\end{eqnarray*}
In other words, $X$ and $Y$  satisfy $X\oplus Y=U\cdot V$.

Using such NL boxes, we can understand Bell's theorem in very intuitive way, since the
{\em CHSH Bell inequality\/} corresponds to an upper bound on the success probability 
of simulating an NL box. 

Let the inputs to the NL box be chosen at random. Let us first look at the classical simulation
of an NL box. It is easy to see that a randomization of the strategies 
does not improve the probability of a correct output. Therefore, both A and B just have
two values---one for input $0$ and one for input $1$---that they will output. Since there
is always a pair of inputs such that the output of the two players is wrong, they can achieve
an accuracy of at most $0.75$. If the players have access to an EPR state $|\psi^-\rangle$,
they can carry out  measurements
in bases that are rotated with respect to each other by the angle of $\pi/8$, and get  an accuracy of
\begin{equation}
\cos^2 \left (\frac \pi 8 \right ) \approx 0.85\ .
\end{equation}

Furthermore, it has been shown that one call to an NL box allows for {\em perfectly\/} simulating
the joint behavior of a singlet state under arbitrary von Neumann measurements~\cite{cgmp}.
It is, therefore, fair to say that the NL box is the classical analogue of a Bell state.


\subsection{Oblivious Transfer}

{\em Oblivious transfer}, introduced by Wiesner~\cite{wiesner} under the name of ``multiplexing''
and  by Rabin~\cite{rabin}, is a primitive of paramount importance in cryptography, in 
particular two- and multi-party computation. In {\em chosen 1-out-of-2 oblivious transfer\/} 
or $\ot$ for short, one party, called the {\em sender}, has two binary inputs $b_0$ and $b_1$,
whereas the other party, the {\em receiver}, inputs a {\em choice bit\/} $c$. The latter then
learns $b_c$ but no additional information, while the sender remains ignorant about $c$.

\begin{figure}[h]
\begin{center}
\input{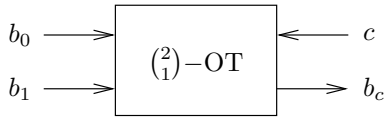}
\end{center}
\caption{Chosen 1-out-of-2  oblivious transfer.}
\label{otbox}
\end{figure}

OT has been shown universal for multi-party computation, i.e., {\em any\/} 
secure computation can be carried out if OT is available. 
An example is  {\em secure function 
evaluation}, which is an important special case of multi-party computation, and where a number of 
players want to secretly and correctly evaluate a function to which each player
holds an input; here, ``secretly'' means that no (unnecessary) information about 
the players' inputs is revealed. 

In this note, we are interested in OT from the viewpoint of communication 
complexity rather than cryptography. In particular, the described reductions 
are not cryptographic: a party can obtain {\em more\/} information than 
specified. 

In~\cite{ww05}, it has been shown that $\ot$ and the NL box are roughly the 
same primitive in a cryptographic sense, i.e., one can be reduced to the 
other. An interesting and  somewhat surprising consequence thereof is 
that $\ot$, just as the NL box, is {\em symmetric\/}: $\ot$ from $A$ to $B$ 
can be perfectly reduced to a single instance of $\ot$ from $B$ to $A$~\cite{otto}. 

In terms of communication complexity, we have the following reductions. 
$\ot$ can be reduced to one realization of an NL box plus one bit 
of (classical) communication as follows. Let $(b_0,b_1)$ and $c$ be
the parties' inputs for $\ot$. Then they input $b_0\oplus b_1$ and $c$ 
to the NL box, respectively, and receive the outputs $X$ and $Y$ with 
$X\oplus Y=(b_0\oplus b_1)c$.  The first party then sends $X\oplus b_0$
to the other, who computes
\[
(X\oplus b_0)\oplus Y=b_0\oplus (b_0\oplus b_1)c=b_c\ ,
\]
as desired. 

Conversely, $\ot$ can be used to realize an NL box. If the inputs to the latter 
are $u$ and $v$, then the parties input 
$(r,r\oplus u)$ (where $r$ is a random bit) and $v$ to the $\ot$. The sender's 
output is then $r$, and the receiver's output is his output from $\ot$, i.e., 
$r\oplus uv$. We, hence, have for the outputs that 
\[
r\oplus (r\oplus uv)=uv\ ,
\]
as requested.

\subsection{Classical Teleportation}

Quantum teleportation~\cite{tele} is the simulation of sending a qubit over a quantum
channel, using   a shared EPR pair and two bits of classical communication. 
{\em Classical teleportation\/}~\cite{ct} is the term used for the following 
reduction. Given a box simulating the classical behavior of an EPR pair, e.g.,
a singlet $|\psi^-\rangle$, and one bit of classical communication, 
one can simulate sending a qubit over a quantum channel and carrying out 
a von Neumann measurement on the received qubit. The idea is as follows~\cite{ct}:
Alice chooses, for her part of the EPR pair,
 a measurement basis consisting of the state she wants to send and its 
orthogonal complement, and communicates the outcome to Bob, who inverts his 
measurement result if and only if Alice has measured the state orthogonal 
to the one to be sent. Together with the result that an NL box allows for 
simulating the behavior of an EPR pair under von Neumann measurements, 
this leads to a realization of classical teleportation with {\em one use of an NL box and one 
bit of classical communication}. When another|weaker|result, by Bacon and Toner~\cite{bt},
is used  stating that the EPR behavior under projective measurements 
can be simulated using one bit of communication, then one obtains the 
possibility of classical teleportation using  two bits of 
classical communication. 

\section{Connection between Oblivious Transfer and Quantum Channels}

In this section we show that---instead of one NL box and one classical bit of communication---one oblivious transfer 
(and shared randomness) is enough to perfectly simulate a quantum channel
with a von Neumann measurement.
Note that our discussion in the previous section shows that 
this is a strictly weaker primitive.

Furthermore, in the case of an NL box and EPR pairs,
the inverse is true as well, with identical success probabilities:
Using a quantum channel, one can simulate oblivious transfer with a probability of about $0.85$,
whereas with a classical channel one cannot be better than $0.75$.

We can, therefore, conclude that oblivious transfer is the classical analogue to a quantum channel,
in the same way as an NL box is for the EPR pair.

\subsection{Simulating a Quantum Channel using Oblivious Transfer}

\begin{figure}[h]
\begin{center}
\input{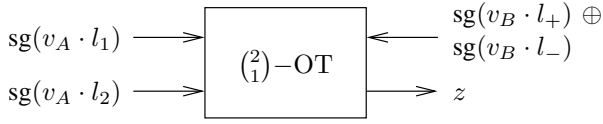}
\end{center}
\caption{Reducing classical teleportation to oblivious transfer.}
\label{red}
\end{figure}

The reduction of classical teleportation to $\ot$ works as follows.
Let $l_1$ and $l_2$ be two random vectors on the Poincar\'e sphere, and 
let $l_{\pm}:=l_1\pm l_2$. (These vectors are the randomness shared
by the two parties.) Let $v_A$ and $v_B$ be the vectors determining 
the state Alice wants to send and the measurement Bob wants to
perform, respectively. Then the inputs to $\ot$ are as shown in Figure~\ref{red}. 
(Here, sg denotes the signum function.) 

Bob's output is  the bit 
\[
z\oplus {\rm sg}(v_B\cdot l_+)\oplus 1\ .
\]
We have 
\[
z={\rm sg}(v_A\cdot l_1)\oplus [{\rm sg}(v_A\cdot l_1)\oplus {\rm sg}(v_A\cdot l_2)][{\rm sg}(v_B\cdot l_+)\oplus {\rm sg}(v_B\cdot l_-)]\ .
\]
Hence, we obtain  the same expression as in~\cite{ct} (see also~\cite{bt}). 

\subsection{Simulating Oblivious Transfer using a Quantum Channel}

Let now Alice and Bob be connected by a quantum channel over which
they are allowed to send exactly one qubit. 
They want to simulate OT (with an error as small as
possible) with Alice as sender and Bob as receiver. Alice has inputs $b_0$ and
$b_1$ and Bob $c$, i.e.,  Bob wants to know $b_c$. Alice takes a qubit $|\phi\rangle = |0\rangle$ and
rotates it by an angle of
\begin{equation}
\phi(b_0,b_1) := \frac \pi 8 (2 b_0 + 4 b_1 - 3)\ .
\end{equation}
She gets
\begin{equation}
|\phi\rangle= \cos(\phi(b_0,b_1))|0\rangle+\sin(\phi(b_0,b_1))|1\rangle
\end{equation}
and send $|\phi\rangle$ to Bob. If $c=1$, Bob applies a Hadamard transform
on $|\phi\rangle$ and leaves it unchanged otherwise. He then measures $|\phi\rangle$ in the computational  basis and outputs
the outcome of this measurement. It is easy to verify that his output is equal to $b_c$ with a
probability of
\begin{equation}
\cos^2 \left (\frac \pi 8 \right ) \approx 0.85\ ,
\end{equation}
which is the same as for the simulation of an NL box using an EPR pair.

If Alice and Bob are only allowed to communicate one classical bit, the best Alice
can do is to choose one of the two input bits and send it to Bob. Therefore, Bob will
only be able to output the correct value $b_c$ with probability $0.75$, if all
inputs are random. Again, 
we get the same success probability 
as for the NL box without an EPR pair.
Results related to ours were obtained in~\cite{antv}. 

\newpage

\section{Concluding Remarks and Open Problems}
We have shown that bit oblivious transfer can be seen as the classical analogue 
of sending a qubit over a quantum channel in the same sense as the NL box is for 
measuring an EPR pair. 

Note that our simulation of a quantum channel does not preserve \emph{privacy}, i.e. the players get more information than
they would get if they had only black-box access to a quantum channel. It is an open problem if there is an efficient
simulation that would also be private.

\section*{Acknowledgment}

This work was supported by the Swiss National Science Foundation (SNF).

\end{document}